\documentclass[letterpaper,twocolumn]{revtex4-1}

\usepackage{amsmath}
\usepackage{xfrac}

\begin{document}

\title{Universal mechanism for the binding of temporal cavity solitons}

\author{Yadong Wang$^1$}
\author{Fran\c{c}ois Leo$^{1,2}$}
\author{Julien Fatome$^{1,3}$}
\author{Miro Erkintalo$^1$}
\author{Stuart~G.~Murdoch$^1$}
\author{St\'ephane Coen$^1$}

\affiliation{$^1$ The Dodd-Walls Centre for Photonic and Quantum Technologies, Physics Department,
                         The University of Auckland, Private~Bag~92019, Auckland~1142, \mbox{New Zealand}}

\affiliation{$^2$ OPERA-Photonique, Universit\'e Libre de Bruxelles, CP~194/5, 50~Av. F.~D. Roosevelt, B-1050
Brussels, Belgium}

\affiliation{$^3$ Laboratoire Interdisciplinaire Carnot de Bourgogne (ICB), UMR 6303 CNRS --- Universit\'e de
            Bourgogne Franche-Comt\'e, 9 Avenue Alain Savary, BP 47870, F-21078 Dijon, France}

\begin{abstract}
  We present theoretical and experimental evidence of a universal mechanism through which temporal cavity solitons
  of externally-driven, passive, Kerr resonators can form robust long-range bound states. These bound states,
  sometime also referred to as multi-soliton states or soliton crystals in microresonators, require perturbations
  to the strict Lugiato-Lefever mean field description of temporal cavity solitons. Binding occurs when the
  perturbation excites a narrowband resonance in the soliton spectrum, which gives long oscillatory tails to the
  solitons. Those tails can then interlock for a discrete set of temporal separations between the solitons. The
  universality of this mechanism is demonstrated in fiber ring cavities by providing experimental observations of
  long-range bound states ensuing from three different perturbations: third-order dispersion (dispersive wave
  generation), the periodic nature of the cavity (Kelly sidebands), and the random birefringence of the resonator.
  Sub-picosecond resolution of bound state separations and their dynamics are obtained by using the dispersive
  Fourier transform technique. Good agreement with theoretical models, including a new vector mean-field model, is
  also reported.
\end{abstract}

\maketitle

\section{Introduction}

Temporal cavity solitons (CSs) are ultrashort light pulses that can persist indefinitely on top of a weak homogeneous
background in a coherently, externally driven, nonlinear passive resonator \cite{leo_temporal_2010}. Their existence
results from a double balance. On the one hand, temporal spreading arising from group velocity dispersion (GVD) is
arrested by a material nonlinearity. On the other hand, the losses they experience through successive passes around
the cavity are compensated for by energy extracted from the coherent driving beam. Temporal CSs have been observed
for the first time in single mode optical fiber ring resonators, where their potential for all-optical buffer
applications has been demonstrated \cite{leo_temporal_2010, jang_temporal_2015, jang_all-optical_2016}. They have
also been shown to underlie the formation of broadband coherent frequency combs in Kerr microresonators
\cite{coen_modeling_2013, herr_temporal_2014, yi_soliton_2015, joshi_thermally_2016, webb_experimental_2016}. Such
microresonators are now foreseen as a promising avenue for on-chip generation of optical frequency combs, with the
potential to enable more widespread use of frequency combs and their applications
\cite{kippenberg_microresonator-based_2011}.

Despite their dramatic difference in scale, both fiber ring resonators and microresonators are very well described by
the same one-dimensional mean-field model, namely the Lugiato-Lefever equation (LLE); temporal CSs correspond to
localized solutions of that equation \cite{lugiato_spatial_1987, haelterman_dissipative_1992, leo_temporal_2010,
matsko_mode-locked_2011, leo_dynamics_2013, coen_modeling_2013, chembo_spatiotemporal_2013}. In the context of the
standard LLE (pure Kerr nonlinearity; GVD truncated at 2nd order), temporal CSs have the form of a single peak
connected to the low-power background by strongly damped oscillatory tails \cite{parra-rivas_dynamics_2014}. Adjacent
CSs do not interact provided their tails do not overlap, which occurs for separations exceeding a few characteristic
soliton widths. Under these conditions, several temporal CSs can co-exist simultaneously in the resonator with
arbitrary separations. The LLE also exhibits translational invariance, associated with the presence of a zero
eigenvalue, otherwise known as a Goldstone mode, in the excitation spectrum of the CSs \cite{firth_optical_1996,
ackemann_chapter_2009}. It implies that perturbations to CS positions are undamped: if some external influence
happens to shift the position of a CS, no restoring force exists to bring the soliton back to its original position.
In the presence of broadband noise, and when multiple temporal CSs are present in the resonator, one could therefore
expect erratic random-walk changes in the separations between the CSs \cite{firth_optical_1996,
parra-rivas_interaction_2017}.

Many experimental observations do not however match that description. Multi-soliton states made up of CSs that are
\emph{widely} separated, reported both in fiber rings and in many microresonator configurations, often appear
\emph{frozen} over very long timescales, sometimes exceeding hours \cite{herr_temporal_2014, jang_bound_2014,
delhaye_phase_2015, brasch_photonic_2016, yi_active_2016, joshi_thermally_2016, lamb_stabilizing_2016,
webb_experimental_2016}. The absence of relative motion between the constituent CSs in these observations suggests
the existence of long-range binding mechanisms that counteract the effect of noise on the Goldstone mode. In this
Article, we describe a universal mechanism that explains the formation of such long-range temporal CS bound states,
and we identify several specific physical processes that lead to this scenario. In particular, we consider in detail,
theoretically and experimentally, the case of CS bound states mediated by dispersive waves (DWs) (or Cherenkov
radiation) emitted in the normal GVD region by temporal CSs perturbed by high-order dispersion
\cite{milian_soliton_2014-1, jang_observation_2014}. This case is particularly relevant in the microresonator
frequency comb context \cite{brasch_photonic_2016}. We then present additional measurements of bound states mediated
by Kelly-like sidebands arising from the periodicity of the cavity, as well as birefringence of the resonator, to
further illustrate the universality of the binding mechanism we propose. In our work, bound states are identified by
measuring their formation dynamics in real time thanks to the so-called dispersive Fourier transform (DFT)
\cite{wetzel_real-time_2012, goda_dispersive_2013, runge_coherence_2013, herink_resolving_2016}. Finally, we also
briefly discuss how other previously reported observations of CS bound states all fit into the general framework that
we describe.

\section{Universal temporal CS binding mechanism}
\label{sec:general}

\begin{figure}[b]
  \centerline{\includegraphics{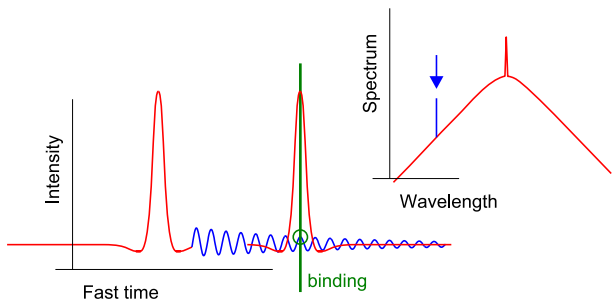}}
  \caption{Schematic illustration of the universal temporal CS binding mechanism. A perturbation leads to a sideband
    in the soliton spectrum (blue arrow). That sideband is associated with an extended oscillatory
    tail in the temporal intensity profile. An adjacent CS gets trapped by the tail oscillations.}
  \label{fig:mechanism}
\end{figure}

We begin by recalling the early 90s theoretical works of Malomed on the formation of soliton bound states
\cite{malomed_bound_1991, malomed_bound_1993}. In these works, Malomed showed that many forms of nonlinear
Schr\"odinger (NLS) equations, with different perturbations, admit a discrete set of long-range soliton bound states.
In every case, bound states arise through excitation, by the perturbation, of a particular frequency in the soliton
spectrum (Malomed, concerned with \emph{spatial} solitons, talked of a particular ``wave number''). That frequency is
associated with an extended oscillatory tail in the soliton envelope, and when tails of adjacent solitons overlap,
this then leads to oscillations in the so-called inter-soliton ``effective interaction potential.'' Each period of
the oscillation constitutes a potential well for a bound state with a specific separation between the solitons.
Essentially, bound states arise through the interlocking of the oscillatory tails of adjacent solitons. Long
interaction ranges are obtained provided the tail extends over many soliton widths, which is associated with a
spectral resonance that is narrow with respect to the soliton bandwidth. In other words, any perturbation that gives
rise to a spectral sideband in the soliton spectrum may lead to the formation of long-range bound states. This
mechanism is illustrated schematically in Fig.~\ref{fig:mechanism}.

The standard LLE, in particular, can be seen as a perturbed NLS equation, with extra damping and AC driving terms.
Previous analyzes have shown however that, in this case, the oscillatory tails are strongly damped for all values of
the cavity detuning but the smallest ones \cite{cai_bound_1994, barashenkov_bifurcation_1998,
parra-rivas_interaction_2017}. It is only when the detuning is less than the width of a cavity resonance that
substantial oscillations exist and that bound states can be observed in practice, as was demonstrated with
\emph{spatial} CSs \cite{schapers_interaction_2000}. For larger values of the detuning, typical of temporal CS
experiments, bound states of this nature have not been reported. In the following, we will show that Malomed's
framework can be extended to perturbed forms of the LLE itself, providing a simple way of estimating long-range CS
bound state separations. This was also recently studied in \cite{parra-rivas_interaction_2017}. In particular, it is
already well-known that different forms of perturbations to the LLE, including high-order dispersion
\cite{milian_soliton_2014-1, jang_observation_2014}, mode-crossings \cite{herr_mode_2014-1}, and acoustic effects
\cite{jang_ultraweak_2013} give CSs extended oscillatory tails. Acoustic-mediated bound states of temporal CSs have
even been observed in optical fiber rings \cite{jang_ultraweak_2013}. These easily fit within our universal
description: the spectral sidebands arise through phase modulation of the CS background by the refractive index
perturbation created in the fiber core by the acoustic wave.

\section{Experimental setup}

With the aim to provide a clear experimental illustration of the universal concepts outlined above, we will present
in the following sections measurements of CS bound states for three different mechanisms. All our experiments are
done in fiber ring resonators and we begin by considering the case of DW-induced long-range temporal CS bound states,
where the perturbation is due to 3rd-order dispersion. Here we detail the experimental setup for this case. The
setups used for the other experiments are very similar: differences will be highlighted in the corresponding
Sections.

\begin{figure}
  \centerline{\includegraphics[width=8.5cm]{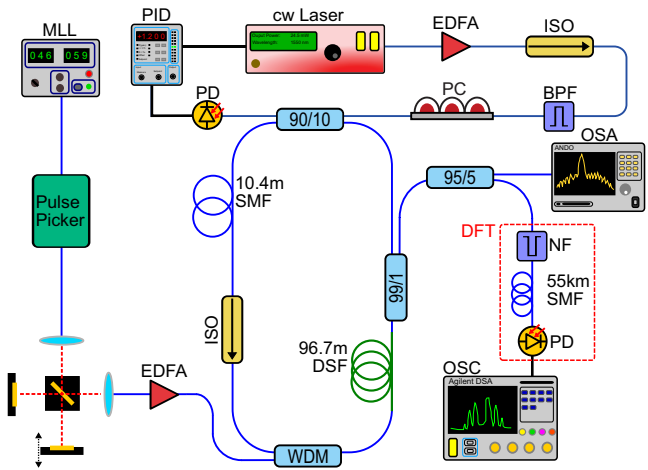}}
  \caption{Experimental setup. PC: polarization controller; PD: photo-diode; other acronyms defined in the text.
    DSF: ${\beta_2 = 1.95}\ \mathrm{ps^2/km}$, ${\beta_3 = 0.2\ \mathrm{ps^3/km}}$,
    ${\gamma = 2/(\mathrm{W\,km})}$; SMF: ${\beta_2 = -21.4\ \mathrm{ps^2/km}}$,
    ${\beta_3 = 0.1\ \mathrm{ps^3/km}}$, ${\gamma = 1.2/(\mathrm{W\,km})}$.}
  \label{fig:setup}
\end{figure}

Our first experiment is performed in a dispersion-managed fiber ring cavity, made up of two different fiber types,
similar to that in \cite{jang_observation_2014} (see setup in Fig.~\ref{fig:setup}). By carefully adjusting the
relative lengths of the two fibers, the average 2nd-order GVD coefficient $\langle\beta_2\rangle$ can be selected so
as to optimize DW emission. In practice, we used about $95.9$~m of dispersion-shifted fiber (DSF) and $10.3$~m of
standard single-mode fiber (SMF). These fibers exhibit, respectively, normal and anomalous GVD at the 1550~nm driving
wavelength, with 2nd and 3rd-order GVD coefficients $\beta_2, \beta_3$ and nonlinearity parameters~$\gamma$ listed in
the caption of Fig.~\ref{fig:setup}. Overall, the average cavity GVD is slightly anomalous at the driving wavelength,
$\langle\beta_2\rangle \simeq -0.31\ \mathrm{ps^2/km}$, with a zero-dispersion wavelength (ZDW) of about 1548~nm. The
driving beam is provided by a narrow-linewidth continuous-wave (cw) laser, amplified up to $1.4$~W with an
Erbium-doped fiber amplifier (EDFA1), and band-pass filtered (BPF) for amplified spontaneous emission (ASE) noise
rejection. It is then coupled into the cavity through a 90/10 fiber coupler. As in \cite{jang_observation_2014}, we
use the light reflected off the cavity by that coupler as a feedback signal to control and lock the detuning of the
driving laser with respect to the cavity resonances. The cavity also includes an optical isolator to avoid stimulated
Brillouin scattering~\cite{leo_temporal_2010}, a 1\,\% tap coupler to monitor the intracavity field, and a
wavelength-division multiplexer (WDM), used for excitation of temporal CSs. The total roundtrip power loss is about
30\,\%, corresponding to a measured cavity finesse of~17.

To study CS bound states and their dynamics, we start each measurement by exciting two temporal CSs in close
proximity using optical addressing pulses~\cite{leo_temporal_2010}. The addressing pulses are generated by an
external mode-locked laser running at a different wavelength (1532~nm) than the cw driving beam. A single pulse is
selected out of the mode-locked pulse train, on request, by a pulse picker. This pulse then passes through an
unbalanced Michelson interferometer to create a delayed replica with a controllable separation and both are
subsequently amplified to about 10~W peak power (EDFA2) before entering the cavity through the WDM. The pair of
addressing pulses circulates for only one roundtrip in the cavity, during which they perturb the cw intracavity field
through cross-phase modulation~\cite{leo_temporal_2010}, before being dumped out by the same WDM. Two temporal CSs
eventually emerge (centered at the driving wavelength) from this perturbation. Once the CSs are excited, EDFA2 is
switched off and the pulse picker is kept blocked so that no other addressing pulses can affect the system.

To measure the separation between the pair of temporal CSs circulating in the cavity, and to monitor in real time the
formation and dynamics of bound states, we send the output light from the tap coupler through 55~km of SMF to perform
a real-time dispersive Fourier transform (DFT)~\cite{wetzel_real-time_2012, goda_dispersive_2013,
runge_coherence_2013, herink_resolving_2016, katarzyna_real-time_2017}. After traveling through that fiber, the
temporal intensity envelope of a CS pair is reshaped into its spectral intensity and this is recorded with a
40~GSample/s digital sampling oscilloscope (OSC). The CS separation is then inferred, roundtrip by roundtrip, from
the spectral interference pattern. The measurement is calibrated by comparing the spectrum of a pair of addressing
pulses measured with a standard optical spectrum analyser (OSA) and that obtained with the DFT. Note that the cw
background on which the pair of temporal CSs sits is not compatible with the DFT scheme. It is thus removed before
the 55~km SMF with an extra notch filter (NF; $0.15$~nm bandwidth centered at the 1550~nm driving wavelength).

\section{Dispersive wave induced CS bound states}

\begin{figure}[!t]
  \centerline{\includegraphics{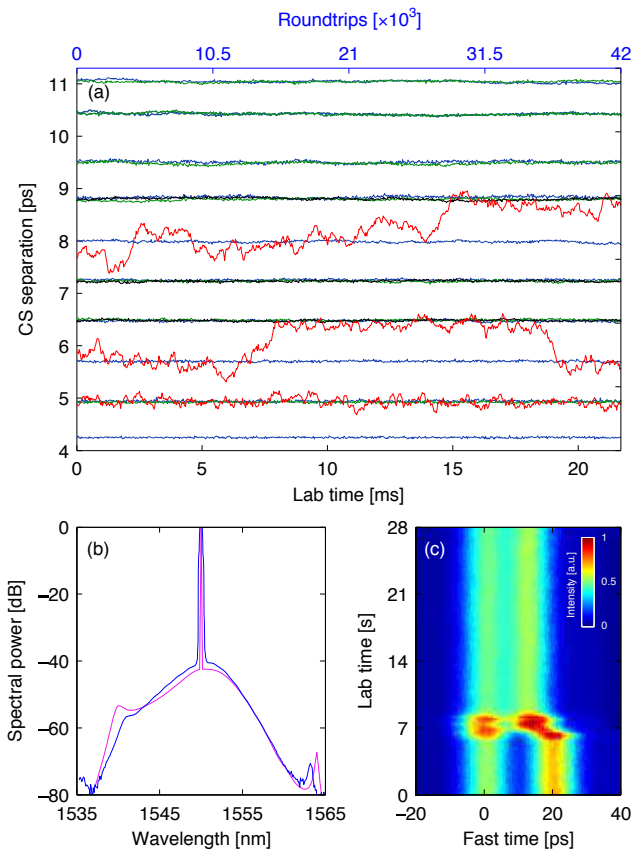}}
  \caption{(a) Dynamical evolution of the separation between pairs of temporal CSs. Overlapping data are shown in
    different colors for clarity (black, green, blue). Red curves are measured in presence of additional ASE noise
    from EDFA2. (b) Measured (blue) and simulated (purple) spectra of an isolated CS. (c) Observation of a transition
    between two bound states when a short burst of ASE noise is injected (at time $t\simeq 7$~s).}
  \label{fig:DWBS}
\end{figure}

Figure~\ref{fig:DWBS}(a) shows results from several DFT measurements in our dispersion-managed cavity. The
measurements were repeated over many independent realizations with a procedure detailed below in order to identify
all possible bound state separations. When the \emph{same} separation was observed multiple times, the corresponding
data are shown in different colors (black, green, and blue curves) in Fig.~\ref{fig:DWBS}(a) for clarity. Each curve
represents an independent recording of the separation of one pair of temporal CSs measured over 42000 roundtrips
(plotted every 60~roundtrips). With a roundtrip time $t_\mathrm{R} \simeq 0.52\ \mu$s, this corresponds to a
relatively large total propagation time (distance) of 22~ms (4500~km), or about 16000 photon lifetimes. It is
immediately apparent that the separation between two temporal CSs cannot take arbitrary values: only a discrete set
of stable separations are observed, ranging in Fig.~\ref{fig:DWBS}(a) from $4.2$~ps to 11~ps. Each of these
corresponds to a particular bound state of temporal CSs. These bound states are very well defined and reproducible:
some of the overlapping measurements have been taken over different days, yet collectively reveal the same set of
stable separations. The observed separations are all much larger (7~to~20 times) than the duration of an individual
temporal CS, which we estimate to be about $0.6$~ps based on the measured spectrum and numerical modeling
[Fig.~\ref{fig:DWBS}(b)]. We can also see in Fig.~\ref{fig:DWBS}(a) that, despite environmental noise, the observed
separations fluctuate by less than 100~fs even over the very long timescale of our measurements. The underlying
binding is thus very robust.

Note that, in our cavity, we have observed bound states with separations of up to 40~ps (or 66 CS width; not shown).
With these larger separations, the bound states are however not as robust: the measured separations fluctuate more,
and the overlap of data taken over successive days is not as good as that shown for smaller separations in
Fig.~\ref{fig:DWBS}(a). In these conditions, we also sometimes witness sudden spontaneous changes from one bound
state separation to another, when environmental noise is strong enough to push one of the solitons into an adjacent
well of the interaction potential. Such spontaneous transitions are never observed in our experiment for separations
below 15~ps. The reduced stability of the larger separations can of course be related to the exponentially decaying
nature of the soliton tails. To demonstrate that these transitions can be caused by noise, we have monitored the
dynamics of bound states in presence of an artificially high level of noise and this is illustrated by the red curves
shown in Fig.~\ref{fig:DWBS}(a). These separation measurements have been obtained with the same DFT technique as the
other curves except that EDFA2 was kept powered on (without an input signal) during the measurements, so that all the
ASE noise it generates is constantly injected into the cavity through the WDM. Because of the particular dispersion
of our cavity, some ASE components are group-velocity matched with the driving beam, guaranteeing efficient noise
transfer. As can be seen, the separation between two CSs vary quite erratically in these conditions, and we can
observe long portions of the data roughly matching bound states measured without noise, with random transitions
between them. A second illustration is provided in Fig.~\ref{fig:DWBS}(c). Here the pseudo-color plot is made up of a
vertical concatenation of oscilloscope recordings of the temporal intensity of the light leaving the cavity over
successive roundtrips (bottom to top). The measurement is performed with a 40~GHz sampling oscilloscope and measured
with a 65~GHz bandwidth photodiode. A bound state of two CSs with a separation of about 20~ps is initially observed
but transitions to a separation of about 12~ps after a short burst of ASE noise is injected into the cavity at time
$t\simeq 7$~s. In fact, it is mostly with this method, applied many times, that we have forced the system to visit
all the separations that bound states of two CSs can possibly adopt, and accumulated the data shown in black, blue,
and green in Fig.~\ref{fig:DWBS}(a). We must also point out here that the influence of EDFA2 ASE noise on bound
states prevent us from directly targeting a specific separation at the excitation stage. Indeed, at that stage, EDFA2
must be powered on to amplify the addressing pulses.

Finally, when looking at all the CS bound states that can be identified in Fig.~\ref{fig:DWBS}(a), we cannot fail to
observe the near periodic arrangement of the allowed separations. This apparent periodicity can be related to
Malomed's picture, in which bound states arise through oscillations in the effective interaction potential. In our
case, these oscillations are mainly due to the excitation of a DW peak at 1541~nm, clearly visible on the CS spectrum
[Fig.~\ref{fig:DWBS}(b)]. That frequency component beats with the cw background on which the temporal CSs sit,
leading to an extended oscillatory tail \cite{milian_soliton_2014-1, jang_observation_2014}. The DW peak is separated
by about $1.13$~THz from the driving beam, corresponding to tail oscillations with a $0.88$~ps period. That latter
figure matches reasonably well with the average $0.75$~ps gap that we observe between allowed separations, thus
strongly suggesting that, in our experiment, binding arises primarily through DWs radiated by the CSs. We must point
out however that the observed separations are actually not perfectly equidistant: for the data shown in
Fig.~\ref{fig:DWBS}(a), the gaps between allowed bound states range from $0.62$~ps to $0.95$~ps and this spread
cannot be attributed to uncertainties in the measurement.

In order to better understand the arrangement of stable separations, we have performed extensive numerical analyzes.
In general, we have found that the possible bound state separations are reasonably well predicted by the position of
the intensity maxima of the oscillatory tail of an isolated CS. This tail appears to be a good approximation of
Malomed's effective interaction potential \cite{malomed_bound_1991, malomed_bound_1993,
parra-rivas_interaction_2017}. We must note that this is also compatible with the well-known fact that CSs are
typically trapped at maxima of a modulated background \cite{firth_optical_1996, jang_temporal_2015}. The oscillatory
tail of the CS can be obtained through numerical modeling of the experiment. In our case, because of the not-so-high
finesse of the fiber resonator, combined with dispersion management, the temporal CSs breathe over each cavity
roundtrip, which breaks the mean-field approximation of the simple LLE \cite{jang_observation_2014}. The simplest
model that qualitatively accounts for all our observations is a version of the LLE with piecewise-constant $\beta_2$,
$\beta_3$, and $\gamma$ coefficients that mimic our two fiber combination. The validity of this approach has been
established in \cite{conforti_modulational_2014}. Note that, for simplicity, the cavity losses are still represented
by a single distributed coefficient, here $\alpha=0.2$. We seek solutions of this model that are invariant (except
for a constant drift) over one full cavity roundtrip using a generalized quasi-Newton (secant) method.

\begin{figure}
  \centerline{\includegraphics{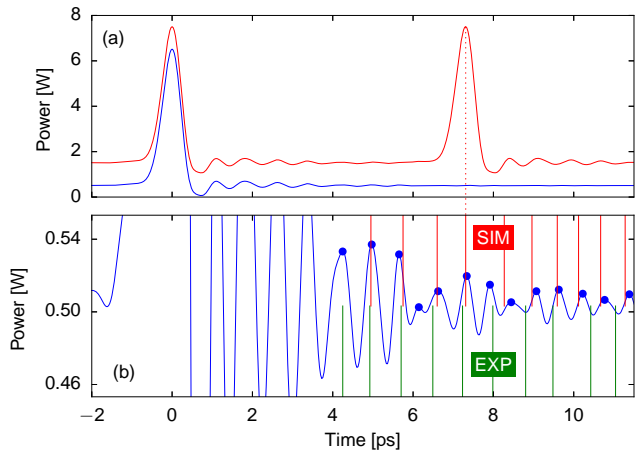}}
  \caption{(a) Simulated temporal intensity profiles of an isolated temporal CS (blue) as well as a stable bound state
    with $7.31$~ps separation (red; shifted vertically for clarity). (b) Vertical enlargement of
    the tail of the isolated CS. Blue dots denote maxima while red (green) lines indicate the separations of simulated
    (experimentally observed) bound states, respectively.}
  \label{fig:sim}
\end{figure}

The blue curve in Fig.~\ref{fig:sim}(a) shows the simulated temporal intensity profile of an isolated CS, obtained
for a driving power of $1.3$~W and a detuning $\delta_0 = 0.57$~rad close to the experimental parameters. The
corresponding spectral profile was depicted in Fig.~\ref{fig:DWBS}(b), and shows good qualitative agreement with
experimental observations. Due to 3rd-order dispersion, the CS tail, shown in more details in Fig.~\ref{fig:sim}(b),
is strongly asymmetric. In~(b), we have highlighted the tail maxima with blue dots and we also show, with red
vertical lines, the stable separations allowed between CSs, which have been obtained by looking for the stationary
two-soliton solutions of our model [one of which is shown in red in panel (a)]. We can generally observe a good
correspondence between the tail maxima and the allowed separations. Differences can be explained by the complexities
of the interactions between the two CSs. First, the main peak of the CS is quite wide in comparison to the period of
the tail oscillations. As a result, trapping is not simply determined by a maximum. Second, on top of interactions
between the main CS peak and the other soliton tail, one must also factor in how the respective tails of the two CSs
interlock with each other, which, given the complex shape of the tail, is not trivial to predict. The complex tail
shape stems in our case from the rather broad bandwidth of the DW, but also from the additional contribution of the
weak Kelly sideband visible at 1563~nm on the spectrum [see Fig.~\ref{fig:DWBS}(b)] \cite{jang_observation_2014,
luo_resonant_2015} (both are due to departure from the mean-field approximation as explained above). These two
spectral features beat with the cw background with non-commensurate frequencies, leading to tail maxima that are not
equidistant and whose amplitude does not decrease monotonically away from the soliton main peak. Some shallow maxima
arise as a result, and as can be seen in Fig.~\ref{fig:sim}(b), these are not necessarily associated with a stable
bound state. Some are also seen to interact with adjacent maxima to affect bound state separations. These mechanisms
all contribute to an aperiodic arrangement of the stable bound state separations, compatible with our experimental
observations. For comparison, the experimentally measured separations are shown in Fig.~\ref{fig:sim}(b) with the
green vertical lines. Given the approximations, and the sensitivity of the model to various parameters, one should
not expect perfect agreement. Still, we observe a qualitatively good match.

\section{Kelly sideband induced CS bound states}

The above results provide very strong support for the hypothesis that the experimentally observed temporal CS bound
states arise due to DW perturbations. To demonstrate more universally that any perturbation that excites spectral
sidebands can lead to distinct bound states, we have performed additional measurements in different cavities that are
dominated by different perturbations. These are presented in this Section and the next.

\begin{figure}[!t]
  \centerline{\includegraphics{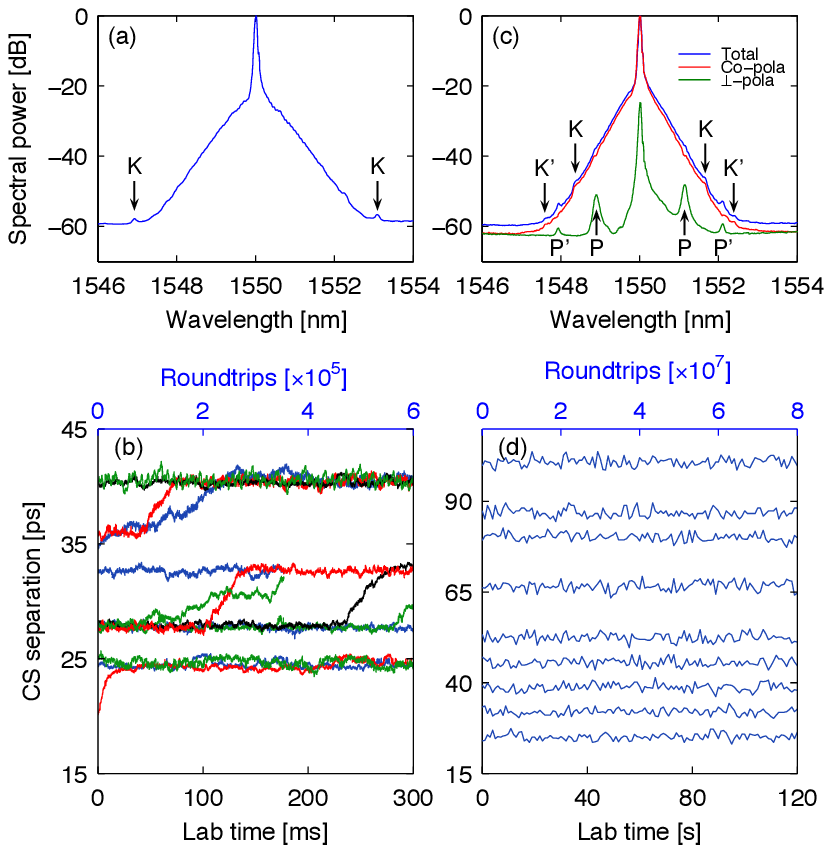}}
  \caption{Spectra of a single temporal CS (top) and dynamical evolution of the separation between
    a pair of such CSs (bottom) in (a), (b) a 100~m-long all-SMF cavity, where CS binding is primarily
    induced by the Kelly sidebands (K), and (c), (d) in a 315~m-long all-SMF cavity,
    where binding arises through excitation of orthogonally polarized resonances
    (P, P'). Notice the different timescales in (b) and~(d) (ms versus~s).}
  \label{fig:KandPola}
\end{figure}

We first consider the case of a fiber cavity of a similar length (100~m) than in our first experiment above but that
is made entirely of SMF, and in which higher-order GVD is negligible. In such a cavity, temporal CSs do not excite
DWs, as confirmed by a spectral measurement [Fig.~\ref{fig:KandPola}(a)]. Yet, very weak Kelly sidebands [labeled K
in Fig.~\ref{fig:KandPola}(a)] are still present $3.1$~nm (or 390~GHz) away from the driving laser. The presence of
the Kelly sidebands can be explained by the limited finesse of the cavity ($21.5$), corresponding to a relatively
large total power loss per roundtrip of about 25\,\%. This loss causes a periodic oscillation (with a period equal to
the cavity roundtrip time) in the power of temporal CSs circulating around the cavity. In turn, this perturbation
leads to the excitation of a quasi-phase matched Kelly sideband in the CS spectrum \cite{gordon_dispersive_1992,
kelly_characteristic_1992, luo_resonant_2015}.

To test for the presence of CS bound states induced by the Kelly sidebands, we have used the same approach as that in
our dispersion-managed cavity. Specifically, we have measured the dynamical evolution of the temporal separation
between two CSs over many realizations and this is reported in Fig.~\ref{fig:KandPola}(b). We still observe a
clustering around a discrete set of values, signalling the presence of bound states, but some important differences
exist. First, the stable (or quasi-stable) separations are observed to be spaced apart by 2--3~ps, much more than in
the dispersion-managed cavity. This is compatible with an interaction potential induced by the Kelly sidebands:
$1/390\ \mathrm{GHz} \simeq 2.6$~ps. We note that Kelly-sideband induced binding has also been reported in
periodically amplified links and in mode-locked fiber lasers \cite{socci_long-range_1999,
soto-crespo_quantized_2003}. Second, it is apparent that the bound states observed in Fig.~\ref{fig:KandPola}(b) are
much more sensitive to noise, with spontaneous transitions occurring in all realizations every fraction of a second,
despite EDFA2 being shut down for these measurements. This can be related to the low amplitude of the Kelly sideband,
which leads to a very shallow interaction potential, in comparison to that due to the high amplitude DW observed in
the dispersion-managed cavity. Overall, these observations are compatible with the universal binding mechanism
introduced in Section~\ref{sec:general}.

\section{Birefringence induced CS bound states}
\label{sec:pola}

We now consider additional measurements performed using a longer (315~m) all-SMF cavity (finesse of 19), with results
shown in Figs.~\ref{fig:KandPola}(c)-(d). Preferred separations, hence bound states, are observed in this cavity too
but with quite different characteristics than in our two previous experiments described above. We can first note that
the bound states revealed by Fig.~\ref{fig:KandPola}(d) are very robust: no spontaneous transitions are observed even
over several minutes. Second, we now observe much larger stable separations, ranging from about 25 to~101~ps. In
fact, because of these larger separations, the measurements plotted in Fig.~\ref{fig:KandPola}(c) have not been
obtained with the DFT technique but directly with an optical sampling scope [the same as that used for
Fig.~\ref{fig:DWBS}(c)]. Finally, we must point out the very regular arrangement of the observed bound states, with a
constant 7~ps gap between allowed separations (excluding a few missing observations). Again this is larger than
reported in the previous Sections, and hints at a different mechanism.

Careful inspection of the temporal CS spectrum for this cavity [blue curve in Fig.~\ref{fig:KandPola}(c)] reveals
several small peaks that could be responsible for the bound states. To gain more insights, we resolved the spectrum
into two orthogonally polarized components. We found that the dominant spectral sidebands (P) are associated with
light that is orthogonally polarized to the driving beam [green curve in Fig.~\ref{fig:KandPola}(c)]. Note how these
peaks do not register on the co-polarized spectrum (red curve). They are shifted by about $\pm 140$~GHz from the
driving frequency (with a slight asymmetry, discussed below) and appear to be responsible for the observed bound
states given that $1/140\ \mathrm{GHz} \simeq 7$~ps matches the gap between allowed separations. To confirm that the
bound states observed in this cavity are related to polarization and birefringence phenomena, we have added a
polarization controller inside our cavity (which is made of standard, non polarization maintaining fiber) to modify
the polarization evolution and the overall cavity birefringence. We found that changing the cavity birefringence led
to a shift in the cross-polarized sidebands P and higher-order sidebands~P', giving rise to a corresponding change in
the spacing between the allowed CS bound state separations. In this cavity, the weaker co-polarized Kelly sidebands,
K and~K' on Fig.~\ref{fig:KandPola}(c), appear to play a negligible role.

To understand better how cross-polarized spectral sidebands can be excited in our cavity, we need to consider in more
details the propagation of light along the fiber. All the cavities discussed in this Article are made of fibers with
some level of random residual birefringence. The polarization state of the intracavity field evolves randomly around
the cavity but there always exist two orthogonal polarization eigenstates for which the polarization state at the end
of the roundtrip matches that at the beginning \cite{coen_experimental_1998}. These polarization eigenstates are
associated with two independent sets of resonances, generally shifted with respect to each other depending on the
overall birefringence. In practice, we always carefully align the driving beam polarization with one of these
polarization eigenstate. However, on top of \emph{linear} (random) birefringence, and in the presence of a temporal
CS, the peak level of the soliton, and the background on which it sits, will experience different levels of
\emph{nonlinear} polarization rotation as they propagate around the cavity. This leads to some light leaking into the
cross-polarized cavity mode. The sidebands P and P' then simply correspond to components of the associated field that
are linearly resonant in the cavity. We note that polarization rotation is essential for this process. In a cavity
made up of a polarization-maintaining fiber, with a driving beam polarized along one of its modes, the field does not
experience any polarization rotation, and therefore no nonlinear polarization rotation either. We have also
identified cross-polarized sidebands in the shorter 100~m-long all-SMF cavity: they are just visible in
Fig.~\ref{fig:KandPola}(a) at an offset of about $2.2$~nm. We infer that the shorter length of that cavity, and of
the dispersion-managed cavity as well, does not provide enough differential nonlinear polarization rotation for the
polarization sidebands to have a notable influence on the bound states.

\begin{figure}[!b]
  \centerline{\includegraphics{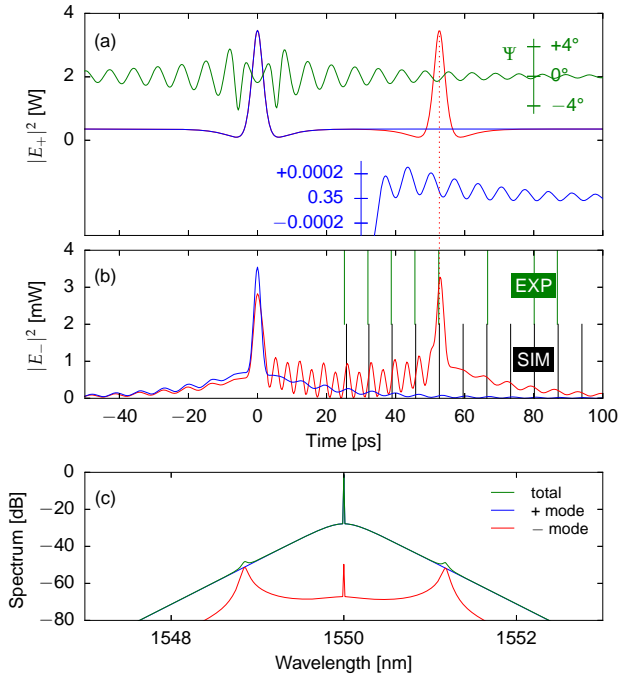}}
  \caption{Simulated temporal intensity profiles of the (a) driven ($+$) and (b) cross-polarized ($-$) modes of
    an isolated temporal CS (blue curves) and that of a particular bound state (red) in a cavity perturbed by fiber
    birefringence. The green curve in (a) is the ellipticity angle $\Psi$ of the field with respect to the driving
    field. (b) also shows simulated (black lines) and measured (green lines) allowed bound-state separations. (c)
    Corresponding spectra.}
  \label{fig:simpola}
\end{figure}

We have confirmed the above description with numerical simulations based on a new vector mean-field model,
Eqs.~(\ref{eq:pola+})--(\ref{eq:pola-}) (see Appendix~A for derivation). Isolated and bound-state CS solutions of
these equations have been obtained with a generalized Newton method. The temporal intensity profile of an isolated
temporal CS is shown as the blue curves in Figs.~\ref{fig:simpola}(a) and (b), respectively for the driven ($+$) mode
and the cross-polarized ($-$) mode. Notice the difference in scales between these two panels (W versus mW),
confirming the perturbative nature of the cross-polarized field. Oscillations with a 7~ps period can be clearly seen
in the tail of the cross-polarized field but are also visible when zooming on the co-polarized field [see sub-axis at
the bottom right of panel (a)]. The dominant spectral sidebands associated with these tail oscillations are
orthogonally polarized to the driving field, as shown on the simulated spectra for the two polarization components
[Fig.~\ref{fig:simpola}(c)]. Notice how these spectra match very well the experimental observations of
Fig.~\ref{fig:KandPola}(c). We have also calculated the ellipticity angle $\Psi$ of the field with respect to the
polarization of the driving field [green curve in panel (a)]. As can be seen, the polarization state varies across
the temporal CS, confirming the role of differential nonlinear polarization rotation between the peak and the
background of the CS. A particular bound-state is then shown as the red curves in panels (a) and~(b), corresponding
to a separation of 53~ps. The black lines in panel (b) further locate all the allowed bound state separations, as
found with the numerical model. They are found to match well the maxima of the tail of the cross-polarized field of
an isolated CS. Panel (b) also shows as green lines the experimental separations, again with a very good agreement
with theoretical simulations.

Overall, our analysis confirms that the bound states observed in our 315~m-long SMF cavity, and illustrated in
Figs.~\ref{fig:KandPola}(c)--(d), result from perturbations involving orthogonally polarized light. To the best of
our knowledge, this is the first experimental observation of a vector form of temporal CSs. The binding mechanism is
also clearly compatible with the universal picture underlying all the other observations reported in this Article.

\section{Conclusion}

In summary, we have studied the bound states of temporal CSs in fiber ring cavities. We have reported experimental
observations of bound states mediated through three different mechanisms involving, respectively, DWs induced by
high-order GVD, Kelly sidebands due to the periodic nature of the cavity, and cross-polarized resonances due to
random fiber birefringence. Despite the differences between these processes, we have highlighted that the resulting
bound states are all underlined by the same universal mechanism: excitation by the CSs of a resonant frequency (or
resonant frequencies) that then lead to long oscillatory tails, and a corresponding long-range oscillating
interaction potential. This can be interpreted as an extension of Malomed's work on perturbed NLS soliton bound
states \cite{malomed_bound_1991, malomed_bound_1993} as has also been described recently in the context of the
generalized LLE \cite{parra-rivas_interaction_2017}. Our measurements clearly highlight that any process generating a
spectral sideband may lead to the formation of long-range bound states between CSs, and that several such processes
can combine together. This conclusion is compatible with previous observations of acoustic-mediated bound states of
temporal CSs \cite{jang_ultraweak_2013} and also with recent suggestions that mode-crossings can sometimes explain
the stability of multi-soliton states in some Kerr microresonators \cite{lamb_stabilizing_2016}. Our observation of
DW-induced bound states would similarly explain the high stability multi-soliton states reported in
\cite{brasch_photonic_2016} and in which mode-crossings are not a dominant feature. Finally, we must point out that
our work illustrates the usefulness of the DFT technique for the dynamical study of soliton bound states (see
also~\cite{katarzyna_real-time_2017}), which could be applicable to other systems in which sub-picosecond resolution
is required.

\section*{Appendix A}

\renewcommand{\theequation}{A\arabic{equation}}

We describe here the model we used to simulate temporal CSs perturbed by the random birefringence of our fiber
cavity, and the corresponding bound states (as described in Section~\ref{sec:pola}). This model underlies the results
shown in Fig.~\ref{fig:simpola}.

To derive our model, we applied the standard mean-field averaging procedure that leads to the scalar LLE
\cite{haelterman_dissipative_1992} to the set of coupled NLS equations that describes elliptically birefringent
fibers as presented in \cite{agrawal_nonlinear_2013} [Eqs.~(6.1.19)--(6.1.20)]. As usual, the cavity finesse is
assumed high enough for the fields not to evolve significantly over one roundtrip, and a corresponding slow time~$t$
is introduced to describe the slow evolution of the fields over subsequent
roundtrips~\cite{haelterman_dissipative_1992}. For our cavity, the two polarization modes described by these
equations will correspond to the polarization eigenstates of the cavity and will be referred to as $+$ and $-$, where
the $+$ mode is the dominant one that is pumped by the external driving laser, whereas the $-$ mode corresponds to
the small amount of cross-polarized light. Because the $-$ mode is a small perturbation, we only keep coherent
coupling terms that are linear in the corresponding field, $E_-$. Overall, this leads to (using the normalization of
\cite{leo_temporal_2010})
\newcommand{\qhalf}{q_{\sfrac{1}{2}}}
\begin{align}
  \frac{\partial E_+}{\partial t} &=
     \left[ -1 + i(|E_+|^2 + B|E_-|^2 - \Delta) -i \eta\frac{\partial^2}{\partial\tau^2} \right] E_+ \nonumber\\
     & \qquad + i D \left[ \qhalf^* E_+^2 E_-^* + 2 \qhalf |E_+|^2 E_- \right] + S\label{eq:pola+}\\
  \frac{\partial E_-}{\partial t} &=
     \left[ -1 + i(|E_-|^2 + B|E_+|^2 - \Delta_\perp)
       -i \eta\frac{\partial^2}{\partial\tau^2} \right. \nonumber\\
     & \qquad \left. - v_\perp \frac{\partial}{\partial\tau} \right] E_-
      + i C q^* E_+^2 E_-^* + i D \qhalf^* |E_+|^2 E_+\label{eq:pola-}
\end{align}
with $E_+ = E_+(t,\tau)$, $E_- = E_-(t,\tau)$. Out of the four coherent coupling terms (with coefficients $C$ and
$D$) present in these equations, the second one in Eq.~(\ref{eq:pola-}) dominates as it only depends on the strong
pumped $E_+$ mode. In fact, that term acts as a \emph{source} term for the cross-polarized field $E_-$. It is the
result of the four-wave mixing product $(+,+) \rightarrow (+,-)$. This source term is essential for our description:
without it, Eq.~(\ref{eq:pola-}) would be trivially satisfied with $E_-=0$ (corresponding to no light in the
cross-polarized mode).

We note that, in Eqs.~(6.1.19)--(6.1.20) of \cite{agrawal_nonlinear_2013}, the two polarization modes are assumed to
keep a fixed elliptical state all along the propagation, whereas the ellipticity and orientation of our polarization
eigenstates evolve along the cavity due to random birefringence. We take this fact into account by averaging the
coupling coefficients $B$, $C$, and $D$ (Eq.~(6.1.21) of \cite{agrawal_nonlinear_2013}) over all possible modal
ellipticities, leading to $B=1.3$, $C=0.2$, $D=0.13$. This approach is different from the usual Manakov approximation
and that corresponds to a full average of polarization coupling terms \cite{agrawal_nonlinear_2013}. In the case of
our cavity, the average is not complete, because the polarization state of the cavity polarization eigenstates are
restored periodically at the end of each roundtrip. This leads to a net leak of power from one mode to the other that
accumulates over subsequent roundtrips, as embodied by the source term of Eq.~(\ref{eq:pola-}) described above. The
incomplete polarization averaging also makes necessary to take into account the difference in group velocities
between the two modes represented by the first-order fast time derivative, with $v_\perp$ the differential group
delay accumulated over one roundtrip.

Other symbols of Eqs.~(\ref{eq:pola+})--(\ref{eq:pola-}) are as follows. As in the usual LLE, $\tau$ is the fast-time
that describes the temporal profile of the fields, $\eta$ is the sign of the GVD coefficient $\beta_2$ ($\eta=-1$
here, corresponding to anomalous dispersion), and $S$ is the driving strength. $\Delta$ and $\Delta_\perp$ measure
the detuning between the driving laser frequency and the closest resonance of, respectively, the driven ($+$) and
cross-polarized ($-$) modes. The two sets of resonances are generally shifted with respect to each other
($\Delta\neq\Delta_\perp$) because of birefringence, and the detunings are related to the difference in cavity
roundtrip phase shift (at the driving frequency) between the two polarization modes, $\phi_{0+}-\phi_{0-} =
\Delta\phi_0 = \alpha(\Delta_\perp - \Delta) + 2k\pi$ (where $\alpha=\pi/\mathcal{F} \simeq 0.165$, with $\mathcal{F}
\simeq 19$ the cavity finesse). $\Delta\phi_0$ is needed to determine the parameters $q = \mathrm{sinc}(\Delta\phi_0)
e^{-i\Delta\phi_0}$ and $\qhalf = \mathrm{sinc}(\Delta\phi_0/2) e^{-i\Delta\phi_0/2}$ that represent the effect of
the phase-mismatch of the various four-wave mixing products. We note that when the two sets of resonances match
($\Delta=\Delta_\perp$), we have $q=\qhalf=0$, and there is no net polarization coupling (even with a multiple of
$2\pi$ phase shift).

The normalized driving power $X=S^2 \simeq 4.7$ and detuning $\Delta \simeq 3$ are determined from corresponding
settings of the experiment. $\Delta_\perp$ and the differential group delay $v_\perp$ can then be found from the
positions of the first order cross-polarized spectral sidebands [$P$ in Fig.~\ref{fig:KandPola}(c)] which satisfy the
resonance condition $\Omega_\mathrm{P}^2 - v_\perp\Omega_\mathrm{P} + \Delta_\perp = 0$ (with $\Omega_\mathrm{P}$ the
angular frequency shift of the sidebands with respect to the driving frequency). We found $\Delta_\perp\simeq -16$
and $v_\perp\simeq -0.1$. Clearly, $v_\perp$ is responsible for the slight asymmetry in the location of the spectral
sidebands. Yet this quantity is very small: it corresponds to a differential group delay of $-75\
\mathrm{fs/roundtrip}$ in real units and indicates that the polarization average is close to complete. We note that
we can only determine $\Delta\phi_0$ to within modulo $2\pi$ with this method. After some trials, we chose an extra
$2\pi$ shift in roundtrip phase ($k=1$) for our simulation as it led to reasonably good agreement with observations.
Larger values of $k$ simply lead to weaker coupling and weaker cross-polarized sidebands.

\section*{Funding Information}

Marsden fund of the Royal Society of New Zealand; Rutherford Discovery Fellowships of the Royal Society of New
Zealand. Julien Fatome also acknowledges the financial support from the European Research Council (ERC starting grant
PETAL \#306633) and the Conseil R\'egional de Bourgogne Franche-Comt\'e (International Mobility Program) which has
allowed him to visit The University of Auckland to contribute to this work.

\section*{Acknowledgements}

We acknowledge experimental support from Jae K. Jang and we thank Lendert Gelens and Pedro Parra-Rivas for numerous
helpful discussions.

\providecommand{\enquote}[1]{``#1''}


\begin{thebibliography}{10}

\bibitem{leo_temporal_2010} F.~Leo, S.~Coen, P.~Kockaert, S.-P. Gorza, {\relax Ph}.~Emplit, and
  M.~Haelterman, \enquote{Temporal cavity solitons in one-dimensional {{Kerr}}
  media as bits in an all-optical buffer,} Nature Photon. \textbf{4}, 471--476
  (2010).

\bibitem{jang_temporal_2015} J.~K. Jang, M.~Erkintalo, S.~Coen, and S.~G. Murdoch, \enquote{Temporal
  tweezing of light through the trapping and manipulation of temporal cavity
  solitons,} Nature Commun. \textbf{6}, 7370 (2015).

\bibitem{jang_all-optical_2016} J.~K. Jang, M.~Erkintalo, J.~Schr{\"o}der, B.~J. Eggleton, S.~G. Murdoch, and
  S.~Coen, \enquote{All-optical buffer based on temporal cavity solitons
  operating at 10\,\,{{Gb}}/s,} Opt. Lett. \textbf{41}, 4526--4529 (2016).

\bibitem{coen_modeling_2013} S.~Coen, H.~G. Randle, T.~Sylvestre, and M.~Erkintalo, \enquote{Modeling of
  octave-spanning {{Kerr}} frequency combs using a generalized mean-field
  {{Lugiato}}-{{Lefever}} model,} Opt. Lett. \textbf{38}, 37--39 (2013).

\bibitem{herr_temporal_2014} T.~Herr, V.~Brasch, J.~D. Jost, C.~Y. Wang, N.~M. Kondratiev, M.~L. Gorodetsky,
  and T.~J. Kippenberg, \enquote{Temporal solitons in optical microresonators,}
  Nature Photon. \textbf{8}, 145--152 (2014).

\bibitem{yi_soliton_2015} X.~Yi, Q.-F. Yang, K.~Y. Yang, M.-G. Suh, and K.~Vahala, \enquote{Soliton
  frequency comb at microwave rates in a high-{{Q}} silica microresonator,}
  Optica \textbf{2}, 1078--1085 (2015).

\bibitem{joshi_thermally_2016} C.~Joshi, J.~K. Jang, K.~Luke, X.~Ji, S.~A. Miller, A.~Klenner, Y.~Okawachi,
  M.~Lipson, and A.~L. Gaeta, \enquote{Thermally controlled comb generation and
  soliton modelocking in microresonators,} Opt. Lett. \textbf{41}, 2565--2568
  (2016).

\bibitem{webb_experimental_2016} K.~E. Webb, M.~Erkintalo, S.~Coen, and S.~G. Murdoch, \enquote{Experimental
  observation of coherent cavity soliton frequency combs in silica
  microspheres,} Opt. Lett. \textbf{41}, 4613--4616 (2016).

\bibitem{kippenberg_microresonator-based_2011} T.~J. Kippenberg, R.~Holzwarth, and S.~A. Diddams,
  \enquote{Microresonator-based optical frequency combs,} Science \textbf{332},
  555--559 (2011).

\bibitem{lugiato_spatial_1987} L.~A. Lugiato and R.~Lefever, \enquote{Spatial dissipative structures in
  passive optical systems,} Phys. Rev. Lett. \textbf{58}, 2209--2211 (1987).

\bibitem{haelterman_dissipative_1992} M.~Haelterman, S.~Trillo, and S.~Wabnitz, \enquote{Dissipative modulation
  instability in a nonlinear dispersive ring cavity,} Opt. Commun. \textbf{91},
  401--407 (1992).

\bibitem{matsko_mode-locked_2011} A.~B. Matsko, A.~A. Savchenkov, W.~Liang, V.~S. Ilchenko, D.~Seidel, and
  L.~Maleki, \enquote{Mode-locked {{Kerr}} frequency combs,} Opt. Lett.
  \textbf{36}, 2845--2847 (2011).

\bibitem{leo_dynamics_2013} F.~Leo, L.~Gelens, {\relax Ph}.~Emplit, M.~Haelterman, and S.~Coen,
  \enquote{Dynamics of one-dimensional {{Kerr}} cavity solitons,} Opt. Express
  \textbf{21}, 9180--9191 (2013).

\bibitem{chembo_spatiotemporal_2013} Y.~K. Chembo and C.~R. Menyuk, \enquote{Spatiotemporal {{Lugiato}}-{{Lefever}}
  formalism for {{Kerr}}-comb generation in whispering-gallery-mode
  resonators,} Phys. Rev. A \textbf{87}, 053852/1--4 (2013).

\bibitem{parra-rivas_dynamics_2014} P.~Parra-Rivas, D.~Gomila, M.~A. Mat{\'\i}as, S.~Coen, and L.~Gelens,
  \enquote{Dynamics of localized and patterned structures in the
  {{Lugiato}}-{{Lefever}} equation determine the stability and shape of optical
  frequency combs,} Phys. Rev. A \textbf{89}, 043813/1--12 (2014).

\bibitem{firth_optical_1996} W.~J. Firth and A.~J. Scroggie, \enquote{Optical bullet holes: Robust
  controllable localized states of a nonlinear cavity,} Phys. Rev. Lett.
  \textbf{76}, 1623--1626 (1996).

\bibitem{ackemann_chapter_2009} T.~Ackemann, W.~J. Firth, and G.-L. Oppo, \enquote{Chapter 6: {{Fundamentals}}
  and applications of spatial dissipative solitons in photonic devices,} Adv.
  At. Mol. Opt. Phys. \textbf{57}, 323--421 (2009).

\bibitem{parra-rivas_interaction_2017} P.~Parra-Rivas, D.~Gomila, P.~Colet, and L.~Gelens, \enquote{Interaction of
  solitons and the formation of bound states in the generalized
  {Lugiato-Lefever} equation,} submitted to Eur. Phys. J.~D  (2017).

\bibitem{jang_bound_2014} J.~K. Jang, M.~Erkintalo, S.~G. Murdoch, and S.~Coen, \enquote{Bound states of
  temporal cavity solitons,} in \enquote{Nonlinear {{Photonics}}, {{NP}}'2014,}
   ({Optical Society of America}, Barcelona, Spain, July 27-31, 2014), OSA
  Technical Digest (online), pp. 2 pages, poster JM5A.59. Accepted:.

\bibitem{delhaye_phase_2015} P.~Del'Haye, A.~Coillet, W.~Loh, K.~Beha, S.~B. Papp, and S.~A. Diddams,
  \enquote{Phase steps and resonator detuning measurements in microresonator
  frequency combs,} Nat Commun \textbf{6}, 5668 (2015).

\bibitem{brasch_photonic_2016} V.~Brasch, M.~Geiselmann, T.~Herr, G.~Lihachev, M.~H.~P. Pfeiffer, M.~L.
  Gorodetsky, and T.~J. Kippenberg, \enquote{Photonic chip\textendash{}based
  optical frequency comb using soliton {{Cherenkov}} radiation,} Science
  \textbf{351}, 357--360 (2016).

\bibitem{yi_active_2016} X.~Yi, Q.-F. Yang, K.~Y. Yang, and K.~Vahala, \enquote{Active capture and
  stabilization of temporal solitons in microresonators,} Opt. Lett.
  \textbf{41}, 2037--2040 (2016).

\bibitem{lamb_stabilizing_2016} E.~S. Lamb, D.~C. Cole, P.~Del'Haye, K.~Y. Yang, K.~J. Vahala, S.~A. Diddams,
  and S.~B. Papp, \enquote{Stabilizing multiple solitons in {{Kerr}}
  microresonator frequency combs,} in \enquote{Conference on {{Lasers}} and
  {{Electro}}-{{Optics}},}  ({Optical Society of America}, 2016), p. SW1E.3.

\bibitem{milian_soliton_2014-1} C.~Mili{\'a}n and D.~V. Skryabin, \enquote{Soliton families and resonant
  radiation in a micro-ring resonator near zero group-velocity dispersion,}
  Opt. Express \textbf{22}, 3732--3739 (2014).

\bibitem{jang_observation_2014} J.~K. Jang, M.~Erkintalo, S.~G. Murdoch, and S.~Coen, \enquote{Observation of
  dispersive wave emission by temporal cavity solitons,} Opt. Lett.
  \textbf{39}, 5503--5506 (2014).

\bibitem{wetzel_real-time_2012} B.~Wetzel, A.~Stefani, L.~Larger, P.~A. Lacourt, J.~M. Merolla, T.~Sylvestre,
  A.~Kudlinski, A.~Mussot, G.~Genty, F.~Dias, and J.~M. Dudley,
  \enquote{Real-time full bandwidth measurement of spectral noise in
  supercontinuum generation,} Sci. Rep. \textbf{2}, 882 (2012).

\bibitem{goda_dispersive_2013} K.~Goda and B.~Jalali, \enquote{Dispersive {{Fourier}} transformation for fast
  continuous single-shot measurements,} Nat Photon \textbf{7}, 102--112 (2013).

\bibitem{runge_coherence_2013} A.~F.~J. Runge, C.~Aguergaray, N.~G.~R. Broderick, and M.~Erkintalo,
  \enquote{Coherence and shot-to-shot spectral fluctuations in noise-like
  ultrafast fiber lasers,} Opt. Lett. \textbf{38}, 4327--4330 (2013).

\bibitem{herink_resolving_2016} G.~Herink, B.~Jalali, C.~Ropers, and D.~R. Solli, \enquote{Resolving the
  build-up of femtosecond mode-locking with single-shot spectroscopy at 90
  {{MHz}} frame rate,} Nat Photon \textbf{10}, 321--326 (2016).

\bibitem{malomed_bound_1991} B.~A. Malomed, \enquote{Bound solitons in the nonlinear
  {{Schr{\"o}dinger}}\textendash{}{{Ginzburg}}-{{Landau}} equation,} Phys. Rev.
  A \textbf{44}, 6954--6957 (1991).

\bibitem{malomed_bound_1993} B.~A. Malomed, \enquote{Bound states of envelope solitons,} Phys. Rev. E
  \textbf{47}, 2874--2880 (1993).

\bibitem{cai_bound_1994} D.~Cai, A.~R. Bishop, N.~Gr{\o}nbech-Jensen, and B.~A. Malomed, \enquote{Bound
  solitons in the {{AC}}-driven, damped nonlinear {{Schr{\"o}dinger}}
  equation,} Phys. Rev. E \textbf{49}, 1677--1679 (1994).

\bibitem{barashenkov_bifurcation_1998} I.~V. Barashenkov, Y.~S. Smirnov, and N.~V. Alexeeva, \enquote{Bifurcation to
  multisoliton complexes in the {{AC}}-driven, damped nonlinear
  {{Schr{\"o}dinger}} equation,} Phys. Rev. E \textbf{57}, 2350--2364 (1998).

\bibitem{schapers_interaction_2000} B.~Sch{\"a}pers, M.~Feldmann, T.~Ackemann, and W.~Lange, \enquote{Interaction
  of localized structures in an optical pattern-forming system,} Phys. Rev.
  Lett. \textbf{85}, 748--751 (2000).

\bibitem{herr_mode_2014-1} T.~Herr, V.~Brasch, D.~Jost, J.\, I.~Mirgorodskiy, G.~Lihachev, L.~Gorodetsky,
  M.\, and J.~Kippenberg, T.\, \enquote{Mode spectrum and temporal soliton
  formation in optical microresonators,} Phys. Rev. Lett. \textbf{113}, 123901
  (2014).

\bibitem{jang_ultraweak_2013} J.~K. Jang, M.~Erkintalo, S.~G. Murdoch, and S.~Coen, \enquote{Ultraweak
  long-range interactions of solitons observed over astronomical distances,}
  Nature Photon. \textbf{7}, 657--663 (2013).

\bibitem{katarzyna_real-time_2017} K.~Katarzyna, K.~Nithyanandan, U.~Andral, P.~Tchofo-Dinda, and P.~Grelu,
  \enquote{Real-time observation of dissipative optical soliton molecular
  motions,} arXiv 1702.01161 (2017).

\bibitem{conforti_modulational_2014} M.~Conforti, A.~Mussot, A.~Kudlinski, and S.~Trillo, \enquote{Modulational
  instability in dispersion oscillating fiber ring cavities,} Opt. Lett.
  \textbf{39}, 4200--4203 (2014).

\bibitem{luo_resonant_2015} K.~Luo, Y.~Xu, M.~Erkintalo, and S.~G. Murdoch, \enquote{Resonant radiation in
  synchronously pumped passive {{Kerr}} cavities,} Opt. Lett. \textbf{40},
  427--430 (2015).

\bibitem{gordon_dispersive_1992} J.~P. Gordon, \enquote{Dispersive perturbations of solitons of the nonlinear
  {{Schr{\"o}dinger}} equation,} J. Opt. Soc. Am. B \textbf{9}, 91 (1992).

\bibitem{kelly_characteristic_1992} S.~M.~J. Kelly, \enquote{Characteristic sideband instability of periodically
  amplified average soliton,} Electron. Lett. \textbf{28}, 806--807 (1992).

\bibitem{socci_long-range_1999} L.~Socci and M.~Romagnoli, \enquote{Long-range soliton interactions in
  periodically amplified fiber links,} J. Opt. Soc. Am. B \textbf{16}, 12--17
  (1999).

\bibitem{soto-crespo_quantized_2003} J.~M. Soto-Crespo, N.~Akhmediev, P.~Grelu, and F.~Belhache, \enquote{Quantized
  separations of phase-locked soliton pairs in fiber lasers,} Opt. Lett.
  \textbf{28}, 1757--1759 (2003).

\bibitem{coen_experimental_1998} S.~Coen, M.~Haelterman, {\relax Ph}.~Emplit, L.~Delage, L.~M. Simohamed, and
  F.~Reynaud, \enquote{Experimental investigation of the dynamics of a
  stabilized nonlinear fiber ring resonator,} J. Opt. Soc. Am. B \textbf{15},
  2283--2293 (1998).

\bibitem{agrawal_nonlinear_2013} G.~P. Agrawal, \emph{Nonlinear Fiber Optics} (Academic Press, 2013), 5th ed.

\end{thebibliography}
\end{document}